\documentclass[useAMS,usenatbib]{mnras}

\usepackage{aas_macros}
\usepackage{amssymb}
\usepackage{booktabs}
\usepackage{graphicx}
\usepackage{multirow}
\usepackage{hyperref}

\usepackage[T1]{fontenc}
\usepackage{aecompl}
\usepackage{mathptmx}

\def\fvar{$F_{\rm var}$\,}

\pagerange{\pageref{firstpage}--\pageref{lastpage}} \pubyear{2015}

\begin{document}

\label{firstpage}

\title[Multifrequency connection]{Locating the $\gamma$-ray emission site in \textit{Fermi}/LAT blazars.\\ \Large II Multifrequency correlations.}

\author[Ramakrishnan et al.]{V.~Ramakrishnan,$^{1}$\thanks{E-mail: venkatessh.ramakrishnan@aalto.fi} T.~Hovatta,$^{1,2}$ M.~Tornikoski,$^{1}$ K.~Nilsson,$^{3}$ E.~Lindfors,$^{4}$
	\newauthor
	M.~Balokovi\'c,$^{2}$ A.~L\"{a}hteenm\"{a}ki,$^{1,5}$ R.~Reinthal$^{4}$ and L.~Takalo$^{4}$\\
	\\
$^{1}$Aalto University Mets\"{a}hovi Radio Observatory, Mets\"{a}hovintie 114, 02540, Kylm\"{a}l\"{a}, Finland\\
$^{2}$Cahill Center for Astronomy and Astrophysics, California Institute of Technology, Pasadena, CA 91125, USA\\
$^{3}$Finnish Centre for Astronomy with ESO (FINCA), University of Turku, V\"{a}is\"{a}l\"{a}ntie 20, 21500 Piikki\"{o}, Finland\\
$^{4}$Tuorla Observatory, Department of Physics and Astronomy, University of Turku, 20100 Turku, Finland\\
$^{5}$Aalto University Department of Radio Science and Engineering, PL 13000, FI-00076 Aalto, Finland}

\maketitle

\begin{abstract}
	In an attempt to constrain and understand the emission mechanism of $\gamma$ rays, we perform a cross-correlation analysis of 15 blazars using light curves in millimetre, optical and $\gamma$ rays. We use discrete correlation function and consider only correlations significant at 99~per cent level. A strong correlation was found between 37 and 95~GHz with a near-zero time delay in most of the sources, and $\sim$1~month or longer in the rest. A similar result was obtained between the optical and $\gamma$-ray bands. Of the 15 sources, less than 50~per cent showed a strong correlation between the millimetre and $\gamma$-ray or millimetre and optical bands. The primary reason for the lack of statistically significant correlation is the absence of a major outburst in the millimetre bands of most of the sources during the 2.5~yr time period investigated in our study. This may indicate that only the long-term variations or large flares are correlated between these bands. The variability of the sources at every waveband was also inspected using fractional rms variability (\fvar). The \fvar displays an increase with frequency reaching its maximum in the $\gamma$ rays.
\end{abstract}


\section{INTRODUCTION}

The rapid variability on various time-scales observed across the electromagnetic spectrum, i.e. from radio to $\gamma$ rays, is one of the characteristic features of a subclass of radio-loud active galactic nuclei (AGNs) called blazars. These are AGNs with relativistic jets directed within a few degrees of our line of sight, thus causing the emission to be Doppler boosted. Spatial localization of the production sites of variable non-thermal continuum emission in blazar jets is crucial for understanding the energy transport in these sources.

Blazars, further sub-classified into BL~Lacertae~objects (BLO) and flat-spectrum radio quasars (FSRQs) show two peaks in the spectral energy distribution (SED). The synchrotron radiation by relativistic electrons in the jet causes the peak in the radio to ultraviolet (and sometimes in X-rays) bands. The high-energy peak is caused by the inverse Compton scattering of seed photons by the electrons causing the synchrotron emission or by ultrarelativistic hadrons from photo-pair production in the jet \citep{Bottcher2013}.

The physical processes associated with the high-energy emission are still uncertain with concerns over the origin of seed photons. Synchrotron self-Compton processes \citep[SSC;][]{Maraschi1992,Bloom1996} are driven by seed photons originating from within the relativistic jet. The external inverse-Compton (EIC) mechanism is responsible when the seed photons originate either from the accretion disk, broad-line region (BLR), and/or the dusty torus \citep[e.g.][]{Dermer1993,Sikora1994,Blazejowski2000}. Correlation analysis is one way of addressing the geometry of the jet, such as the size and structure, and location of emission in blazars.

Ever since the launch of the \textit{Fermi Gamma-ray Space Telescope} in 2008, numerous $\gamma$-ray bright blazars are being detected by its primary scientific instrument, the Large Area Telescope (hereafter \textit{Fermi}/LAT) \citep{Abdo2009}. Dedicated monitoring campaign in support of the \textit{Fermi}/LAT observations in $R$-band led to the optical and $\gamma$-ray correlation analysis by \citet{CohenD2014}. From the analysis of 40 blazars over a time span of 5~yrs, the authors showed the variations in the two bands to be consistent with a zero-time delay. At the radio frequencies, \citet{Fuhrmann2014} studied the connection between the radio (from 2.45--345~GHz) and $\gamma$-ray light curves of 54 blazars using 3.5~yr long data, while \citet{MaxMoerbeck2014a} analysed 4~yr worth of 15~GHz radio and $\gamma$-ray data. In the former, a time delay in the range 7--76~d (in source frame) was reported with the radio (progressing from higher to lower frequencies) lagging the $\gamma$ rays. Similar conclusion was reported in the latter with three sources having time delays: 40, 120 and 150~d. In \citet[hereafter Paper~I]{Ramakrishnan2015}, we addressed the correlations between the 37~GHz radio and GeV $\gamma$-ray light curves for a sample of 55 sources over a time span of 5~yrs. Based on the size of the radio core from VLBA observations and the average time delay, we were able to constrain the $\gamma$-ray emission region to the parsec-scale jet.

A comparison of variations in different wavebands can determine how emitting regions are spatially related. In this paper, we attempt to address some of the questions mentioned above from the cross-correlation of light curves at 37~GHz, 95~GHz, \textit{R}-band and $\gamma$ rays.

In Section~\ref{sect2} we describe the sample, followed by the observation and data reduction methods in Section~\ref{sect3}. The statistical tests such as Power Spectral Density, cross-correlations, and fractional variability are mentioned in Sections~\ref{sect4}--\ref{sect6}. The results are discussed in Section~\ref{sect7} and a summary of current findings in Section~\ref{sect8}.

\section{SAMPLE DESCRIPTION}
\label{sect2}

We use a sample of 15 objects selected for the Monitoring of $\gamma$-ray Active galactic nuclei with Radio, Millimetre and Optical Telescopes (MARMOT) project\footnote{\url{http://www.astro.caltech.edu/marmot/}}. These are sources from a combination of flux-limited samples in $\gamma$ rays and optical so that despite the modest sample size, it will be possible to draw statistical conclusions based on the results. We have first selected the 30 brightest sources from the Second LAT AGN catalog \citep{Ackermann2011} and of those we have kept all sources that have \textit{R}-band magnitude $< 17$. This filter results in a $\gamma$-ray and an optically selected sample of 15 sources that also turn out to be bright at 95~GHz. Our sample includes 12 FSRQs and 3 BLOs, while optical data are available for nine of them.

\section{OBSERVATIONS AND DATA REDUCTION}
\label{sect3}

\subsection{\textit{Fermi}/LAT}
The $\gamma$-ray fluxes for the energy range 0.1--200~GeV were obtained by analysing the \textit{Fermi}/LAT data from 2012 August 1 to 2014 November 15 using the \textit{Fermi} Science Tools\footnote{\url{http://fermi.gsfc.nasa.gov/ssc/data/analysis/documentation/Cicerone}} version v9r33p0. Only events in the source class were selected while excluding events with zenith angle $>100^{\circ}$ to avoid contamination from the photons coming from the Earth's limb. The photons were extracted from a circular region centred on the source, within a radius of 15$^{\circ}$. The instrument response functions P7REP\_SOURCE\_V15 were used.

We implemented an unbinned likelihood analysis using \textit{gtlike} \citep{Cash1979, Mattox1996}, following the same steps as in Paper~I. The Galactic diffuse emission and the isotropic background were modelled using the templates, `gll\_iem\_v05\_rev1.fit' and `iso\_source\_v05\_rev1.txt' provided with the Science Tools. We obtained fluxes from weekly integrations with a detection criterion such that the maximum-likelihood test statistic \citep[TS;][]{Mattox1996} exceeds four ($\sim 2\sigma$), while discarding those bins with TS $<4$ or when the predicted number of photons was less than four.

\subsection{Optical}

In the optical we obtained the \textit{R}-band data for nine sources from the Tuorla blazar monitoring programme\footnote{\url{http://users.utu.fi/kani/1m/index.html}}, Steward Observatory of the University of Arizona\footnote{\url{http://james.as.arizona.edu/~psmith/Fermi/}} (observations under the ground-based observational support of the \textit{Fermi Gamma-ray Space Telescope}) and the Yale University SMARTS\footnote{\url{www.astro.yale.edu/smarts/glast/home.php}} programme. All the optical data were converted from magnitudes to flux densities using the zero-point flux of 3080, 2941 and 3064~Jy for Tuorla, Steward and SMARTS observations.

\subsection{Millimetre}

\subsubsection{CARMA}

CARMA observations were made using the eight 3.5~m telescopes of the array with a central frequency of 95\,GHz and a bandwidth of 7.5\,GHz. The data were reduced using the MIRIAD (Multichannel Image Reconstruction, Image Analysis and Display) software \citep{Sault1995}, including standard bandpass calibration on a bright quasar.  The amplitude and phase gain calibration was done by self-calibrating on the target source. The absolute flux calibration was determined from a  temporally nearby observation (within a day) of the planets Mars, Neptune or Uranus, whenever possible. Otherwise the sources 3C~273, 3C~345, and 3C~84 were used as secondary calibrators. The estimated absolute calibration uncertainty of 10 per cent is added to the error bars of objects with flux density $>1$\,Jy. For the fainter objects, this overestimates the uncertainties and we have only used the statistical uncertainties in the analysis.

\subsubsection{Mets\"{a}hovi}
The 37~GHz observations were obtained with the 13.7~m diameter Mets\"{a}hovi radio telescope, which is a radome enclosed paraboloid antenna situated in Finland. The flux density scale is set by observations of DR~21. Sources NGC~7027, 3C~274, and 3C~84 are used as secondary calibrators. A detailed description of the data reduction and analysis is given in \citet{Terasranta1998}. The error estimate in the flux density includes the contribution from the measurement rms and the uncertainty of the absolute calibration. Data points with a signal-to-noise ratio $< 4$ are handled as non-detections and discarded from the analysis.

Hereafter, the term millimetre corresponds to both 37~GHz and 95~GHz light curve.

\begin{table*}
	\begin{minipage}{180mm}
	\caption{Source sample and PSD results}
	\label{tab1}
	\begin{tabular}{lccccccccc}
	\toprule
	Source & 2FGL Name & Optical class & $z$ & $\alpha_{\rm 95~GHz}$ & $\alpha_{R-{\rm band}}$ & $F_{\rm var, 37~GHz}$ & $F_{\rm var, 95~GHz}$ & $F_{\rm var, \textit{R}-band}$ & $F_{\rm var, \gamma-ray}$ \\
	(1) & (2) & (3) & (4) & (5) & (6) & (7) & (8) & (9) & (10) \\
	\midrule
	0234+285 & J0237.8+2846 & FSRQ & 1.206 & 1.93 [1.78, 2.02] & - & 0.13$\pm$0.01 & 0.16$\pm$0.01 & - & 0.49$\pm$0.01 \\
	0235+164 & J0238.7+1637 & BLO  & 0.940 & 2.32 [2.08, 2.35] & 1.75 [1.54, 1.89] & 0.22$\pm$0.01 & 0.22$\pm$0.01 & 0.62$\pm$0.01 & 0.68$\pm$0.02 \\
	0716+714 & J0721.9+7120 & BLO  & 0.310 & 2.52 [2.33, 2.55] & 1.65 [1.54, 1.87] & 0.40$\pm$0.01 & 0.45$\pm$0.01 & 0.57$\pm$0.01 & 0.67$\pm$0.01 \\
	0736+017 & J0739.2+0138 & FSRQ & 0.189 & 2.47 [2.26, 2.49] & - & 0.19$\pm$0.01 & 0.16$\pm$0.01 & - & 0.53$\pm$0.02 \\
	0805-077 & J0808.2-0750 & FSRQ & 1.837 & 2.31 [2.13, 2.34] & - & 0.29$\pm$0.01 & 0.20$\pm$0.01 & - & 0.61$\pm$0.02 \\
	0917+449 & J0920.9+4441 & FSRQ & 2.188 & 1.28 [1.19, 1.41] & - & 0.19$\pm$0.01 & 0.13$\pm$0.01 & - & 0.01$\pm$0.30 \\
	1222+216 & J1224.9+2122 & FSRQ & 0.434 & 1.63 [1.62, 1.84] & 2.00 [1.69, 2.05] & 0.13$\pm$0.02 & 0.17$\pm$0.01 & 0.54$\pm$0.01 & 0.75$\pm$0.01 \\
	3C~273   & J1229.1+0202 & FSRQ & 0.158 & 2.10 [1.91, 2.17] & 2.25 [2.23, 2.48] & 0.09$\pm$0.01 & 0.15$\pm$0.01 & 0.05$\pm$0.01 & 0.53$\pm$0.02 \\
	3C~279   & J1256.1-0547 & FSRQ & 0.536 & 1.12 [0.95, 1.18] & 1.64 [1.61, 1.83] & 0.13$\pm$0.01 & 0.16$\pm$0.01 & 0.38$\pm$0.01 & 1.20$\pm$0.01 \\
	1510-089 & J1512.8-0906 & FSRQ & 0.360 & 2.21 [2.07, 2.31] & 1.32 [1.13, 1.39] & 0.18$\pm$0.01 & 0.19$\pm$0.01 & 0.25$\pm$0.01 & 0.68$\pm$0.01 \\
	2141+175 & J2143.5+1743 & FSRQ & 0.211 & 1.25 [1.23, 1.41] & - & 0.17$\pm$0.01 & 0.16$\pm$0.02 & - & 0.32$\pm$0.03 \\
	BL~Lac   & J2202.8+4216 & BLO  & 0.068 & 2.03 [2.02, 2.25] & 0.99 [0.76, 1.07] & 0.33$\pm$0.01 & 0.44$\pm$0.01 & 0.58$\pm$0.01 & 0.70$\pm$0.01 \\
	2201+171 & J2203.4+1726 & FSRQ & 1.076 & 1.97 [1.82, 2.01] & - & 0.20$\pm$0.01 & 0.08$\pm$0.02 & - & 0.21$\pm$0.18 \\
	2230+114 & J2232.4+1143 & FSRQ & 1.037 & 2.20 [2.14, 2.32] & - & 0.19$\pm$0.01 & 0.37$\pm$0.01 & 0.48$\pm$0.01 & 0.97$\pm$0.01 \\
	3C~454.3 & J2253.9+1609 & FSRQ & 0.859 & 2.41 [2.23, 2.43] & 1.49 [1.35, 1.60] & 0.58$\pm$0.01 & 0.55$\pm$0.01 & 0.86$\pm$0.01 & 1.24$\pm$0.01 \\
	\bottomrule
	\end{tabular}
	{\footnotesize Columns are: (1) Source name; (2) 2FGL name; (3) optical classification; (4) redshift \citep{Nolan2012}; (5)(6) the first value is the best-fitting PSD slope obtained from Monte Carlo simulations at 95~GHz and \textit{R}-band along with the 68~per cent confidence intervals in square brackets; (7)--(10) fractional rms variability in 37~GHz, 95~GHz, \textit{R}-band and $\gamma$ rays.}
	\end{minipage}
\end{table*}
\begin{figure*}
	\centering
	\includegraphics[width=1.98\columnwidth]{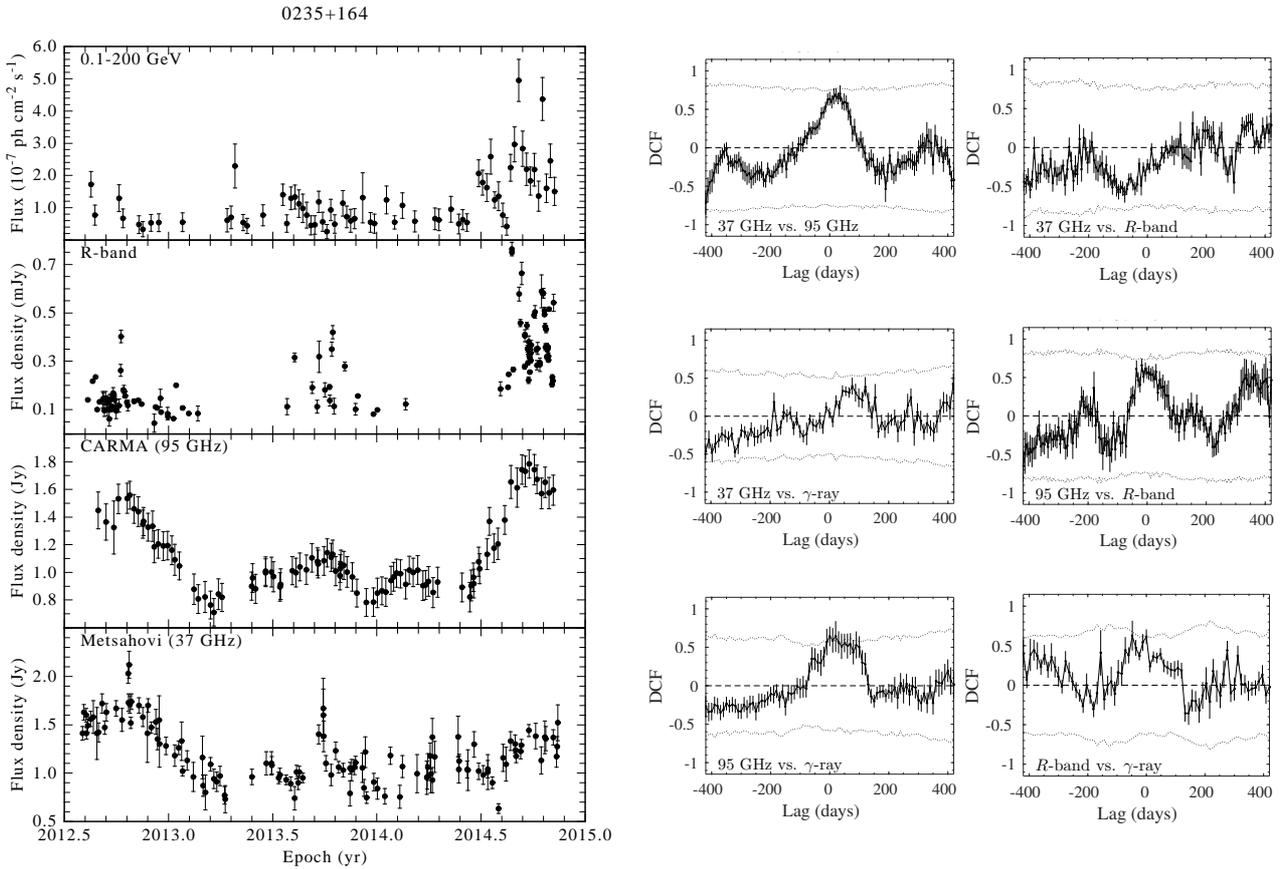}
	\caption{Left: Multifrequency light curve of BLO 0235+164. Right: Cross-correlations for all the wavebands. The dotted line in the correlations corresponds to the 99~per cent significance level estimated from simulations.\hspace{\textwidth} (The complete set of light curves along with the correlations is available from the online journal.)}
	\label{LC_0235}
\end{figure*}

\section{POWER SPECTRAL DENSITY}
\label{sect4}

The Power Spectral Density (PSD) slopes for the sources at all wavebands were estimated following a variant of the Power Spectral Response method \citep{Uttley2002}. The PSD slope of a light curve can be estimated by fitting the periodogram with an underlying model; a power-law (PSD $\propto f^{-\alpha}$) is assumed in this case. However, the finite length (``red-noise leak'') and uneven sampling (``aliasing'') of the light curve introduces a bias in the PSD estimation. To handle these complications, we simulate 1000 light curves following the method proposed by \citet{Emmanoulopoulos2013}. The effect of red-noise leakage and aliasing was taken into account by simulating longer light curves with increased time resolution and through the convolution of Hanning window function with the original light curve \citep{MaxMoerbeck2014b}. Before resampling the simulated light curve to those of the original ones, we added Gaussian noise with variance matching those of the observations. The binned logarithmic periodogram was then computed following \citet{Papadakis1993} and \citet{Vaughan2005} for the observed and all of the simulated light curves.

The goodness of fit is then estimated based on the comparison of average and standard deviation of the simulated periodograms with the periodogram of the observed data. The goodness of fit obtained was tested for a range of PSD slopes. The best-fitting PSD slope corresponds to the one with the highest goodness of fit. For more details on the method see Paper~I, and the references therein.

In Paper~I we estimated the PSD slopes of 51 and 48 sources at 37~GHz and $\gamma$ rays, respectively. We use these estimates for all sources, except 2141+175 which was not included in the sample in Paper~I. For this source, we obtained 1.70 [1.51, 1.87] at 37~GHz and 0.91 [0.81, 1.04] at $\gamma$ rays, respectively.

The estimated PSD slopes are shown in Table~\ref{tab1}. The average PSD slopes at 95~GHz for FSRQs and BLOs are 1.9 and 2.3, respectively. These are comparable to the mean values of 2.2 and 2.0 obtained for FSRQs and BLOs at 37~GHz in Paper~I.

In all optical light curves there are seasonal gaps. Hence, we only considered continuous light curve segments for the PSD analysis, ignoring the gaps. Each light curve had on average two segments. Each segment is treated separately for most of the PSD simulation, similar to those performed at the millimetre band. However, while estimating the goodness of fit towards the end of the simulations, the periodograms estimated from the two segments were brought together. This way we probe all variability time-scales of the data inherent within the given time interval. The average PSD slopes for FSRQs and BLOs are 1.7 and 1.5, respectively. These values are in good agreement with the best-fitting slope, 1.6$\pm$0.3 of average PSD in \textit{R}-band reported in \citet{Chatterjee2012}.

\begin{table}
	\caption{Cross-correlation results for sources with significance $>99$~per cent. A positive time delay implies that the lower frequency lags those at higher freqeuncy and the vice versa for a negative time lag. The 68.27~per cent confidence interval for the time lags are shown within square brackets. Sources with no optical data are marked by asterisk.}
	\label{tab2}
	\begin{tabular}{lcccc}
	\toprule
	& \multicolumn{2}{c}{37~GHz vs. 95~GHz} & \multicolumn{2}{c}{37~GHz vs. \textit{R}-band} \\
	\cmidrule(lr){2-3} \cmidrule(lr){4-5}
	Source & Lag & DCF & Lag & DCF \\
	& (d) &  & (d) &  \\
	\midrule
	0234+285 & 6 [2, 9] & 0.791 & * & * \\
	0716+714 & 3 [1, 5] & 0.910 & - & - \\
	0736+017 & 9 [5, 13] & 0.715 & * & * \\
	0917+449 & $-$8 [$-$22, 6] & 0.791 & * & * \\
	1222+216 & $-$6 [$-$9, $-$2] & 0.771 & - & - \\
	3C~273 & 79 [74, 83] & 0.885 & - & - \\
	1510$-$089 & 5 [1, 8] & 0.786 & - & - \\
	2141+175 & 65 [12, 118] & 0.653 & * & * \\
	BL~Lac & 7 [5, 9] & 0.814 & - & - \\
	2201+171 & 12 [1, 23] & 0.772 & * & * \\
	2230+114 & 51 [45, 56] & 0.910 & 117 [95, 138] & 0.676 \\
	3C~454.3 & - & - & 272 [267, 276] & 0.618 \\
	\midrule
	& \multicolumn{2}{c}{37~GHz vs. $\gamma$-ray} & \multicolumn{2}{c}{95~GHz vs. \textit{R}-band} \\
	\cmidrule(lr){2-3} \cmidrule(lr){4-5}
	0716+714 & - & - & 78 [75, 82] & 0.755 \\
	1222+216 & - & - & 261 [242, 281] & 0.754 \\
	2141+175 & 336 [320, 343] & 0.629 & * & * \\
	2230+114 & 74 [67, 81] & 0.505 & 58 [48, 69] & 0.699 \\
	\midrule
	& \multicolumn{2}{c}{95~GHz vs. $\gamma$-ray} & \multicolumn{2}{c}{\textit{R}-band vs. $\gamma$-ray} \\
	\cmidrule(lr){2-3} \cmidrule(lr){4-5}
	0235+164 & 27 [15, 39] & 0.642 & $-$28 [$-$60, 4] & 0.626 \\
	0716+714 & - & - & 44 [31, 57] & 0.570 \\
	0805-077 & 70 [61, 79] & 0.536 & * & * \\
	1222+216 & - & - & 47 [36, 57] & 0.780 \\
	1510-089 & - & - & 0 [$-$3, 3] & 0.542 \\
	2141+175 & 181 [177, 185] & 0.465 & * & * \\
	BL~Lac & - & - & 29 [16, 43] & 0.425 \\
	2230+114 & 41 [29, 53] & 0.492 & - & - \\
	3C~454.3 & - & - & 0 [$-$3, 4] & 0.868 \\
	\bottomrule
	\end{tabular}
\end{table}

\section{DISCRETE CORRELATION FUNCTION}
\label{sect:dcf}

To study the correlated variability between all the frequencies considered in this work, we used the widely employed discrete correlation function (DCF) of \citet{Edelson1988}. Unlike the classical correlation function by \citet{Oppenheim1975}, the DCF works very well even in the case of unevenly sampled light curves. We applied the local normalization to the DCF, which restricts the DCF peak within [$-$1, +1] interval. The uncertainties of the DCF were estimated based on a model-independent Monte Carlo method proposed by \citet{Peterson1998}. For more discussion on the cross-correlation, see \citet{MaxMoerbeck2014b} and Paper~I.

\begin{figure*}
	\centering
	\includegraphics[width=1.8\columnwidth]{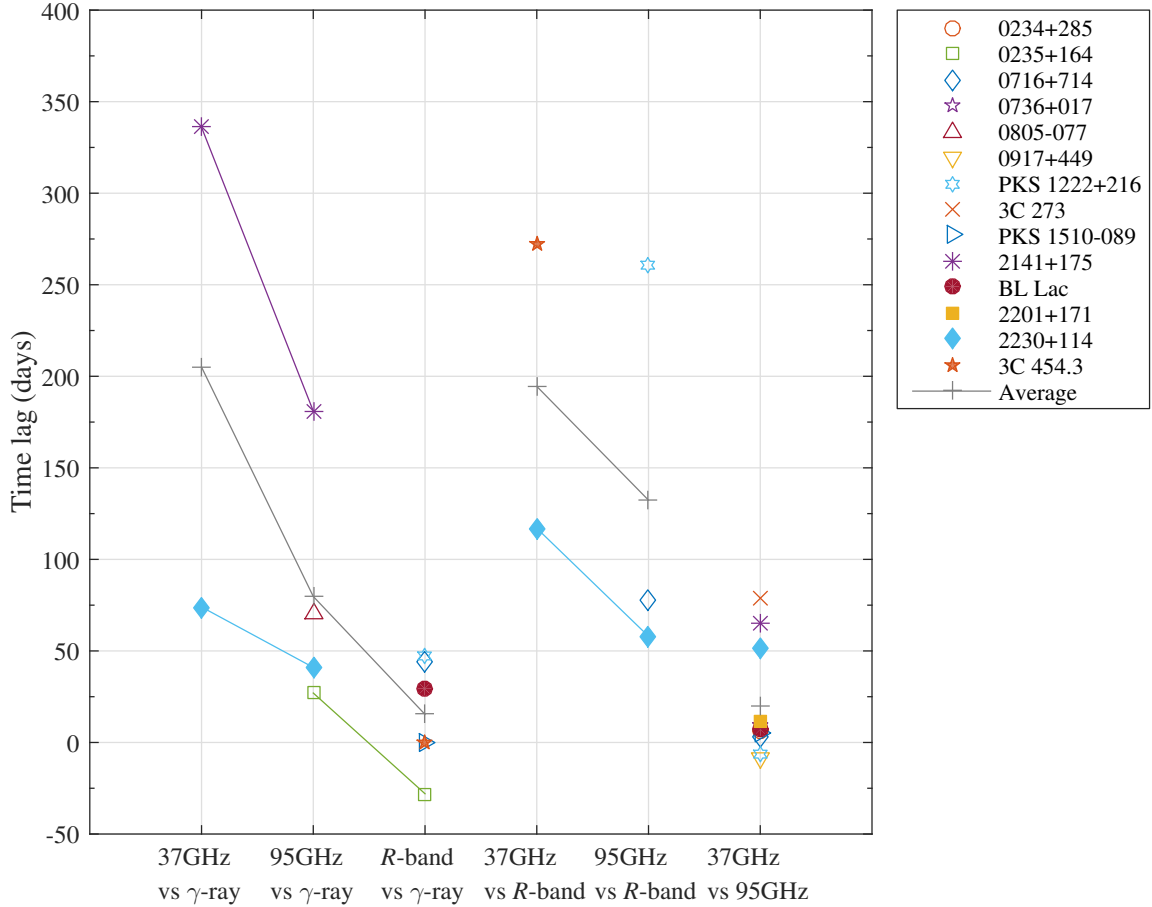}
	\caption{Time lags from cross-correlations reported in Table~\ref{tab2}. The cross-correlations of millimetre and \textit{R}-band with $\gamma$ rays, and millimetre with \textit{R}-band are connected by lines. The average shown in grey plus symbol is the mean estimated from the time lags in the respective cross-correlation analysis. Due to synchrotron self-absorption effects, the time lags increase with decreasing frequency.}
	\label{fig1}
\end{figure*}

An important step after estimating the cross-correlation is to determine the significance of the DCF peak. In this work we adopt the same approach as in Paper~I of cross-correlating simulated light curves to estimate the significance. Using the best-fitting PSD slopes at 95~GHz and \textit{R}-band from Table~\ref{tab1}, and those at 37~GHz and $\gamma$ rays from Paper I, we simulated 1000 light curves at every waveband using the method of \citet{Emmanoulopoulos2013}. The light curves are simulated assuming that the variability properties exhibited by the original light curves are distributed as a power-law. Finally, Poisson noise for $\gamma$ rays using equation~3 in Paper~I and Gaussian noise for simulated light curves at other wavebands were also added. Every simulated light curve was in turn cross-correlated in a similar way to those of the observed light curves. At every time lag, we obtained a cross-correlation distribution from which the 99~per cent significance levels were determined. The time lags and the DCF peak were determined by fitting a Gaussian function. It is important to note that the significance test performed is to probe the probability of obtaining a correlation at a particular time lag; here we look at the time lag associated with the DCF peak. From the cross-correlation analysis only those peaks that are significant at or above 99~per cent level are showed in Table~\ref{tab2}. We show the cross-correlation results along with the multifrequency light curve of the source 0235+164 in Fig.~\ref{LC_0235}.

The time lags obtained from the correlations are plotted in Fig.~\ref{fig1}. The average shown in this plot corresponds to the mean of the time lags for all significant cross-correlations. It is evident that the time lag of correlations with respect to $\gamma$ rays and \textit{R}-band decreases with increasing frequency. This trend is due to synchrotron self-absorption, which causes the maximum of a flare to be first detectable at the highest frequencies, with the peak shifting to lower frequencies when the flare becomes optically thin \citep{Valtaoja1992a}. This frequency dependence on the time lag can also be found in \citet{Fuhrmann2014}. By cross-correlating blazars at radio frequencies (2.64--345~GHz) with those at $\gamma$ rays, the authors showed that the decrease in the time lag with increasing frequency followed a power-law relation.

The sampling of the $\gamma$-ray light curves are affected due to bins with TS < 4 being discarded from the cross-correlation analysis. In the worst case, which is for source 2201+171, we loose $\sim$75~per cent of the data due to this constraint. To this end, we tested the effect of including all the $\gamma$-ray data without imposing the constraint, and by also replacing those bins with TS < 4 with 2$\sigma$ upper limits computed using the profile likelihood method \citep{Rolke2005}. We found no significant difference in the correlation result obtained this way in comparison to those results obtained with the discarded data points. A similar finding was also reported in the work by \citet{MaxMoerbeck2014a}. Because the optical light curves are affected by seasonal gaps, it may also have an effect on the significance of the correlations due to uncertainties in the obtained PSD slopes. To test this effect, we estimated the significance of the correlation by changing the PSD slope in steps of $\pm$~0.1 compared to the best-fitting value in Table~\ref{tab1}. We find that only when the PSD slope differs by more than $\pm$~0.4 do we see a change in the significance of the correlations with the optical band. This shows that our results are fairly insensitive to large changes in the PSD slopes.

\section{Fractional Variability}
\label{sect6}

\begin{figure*}
	\centering
	\includegraphics[width=\columnwidth]{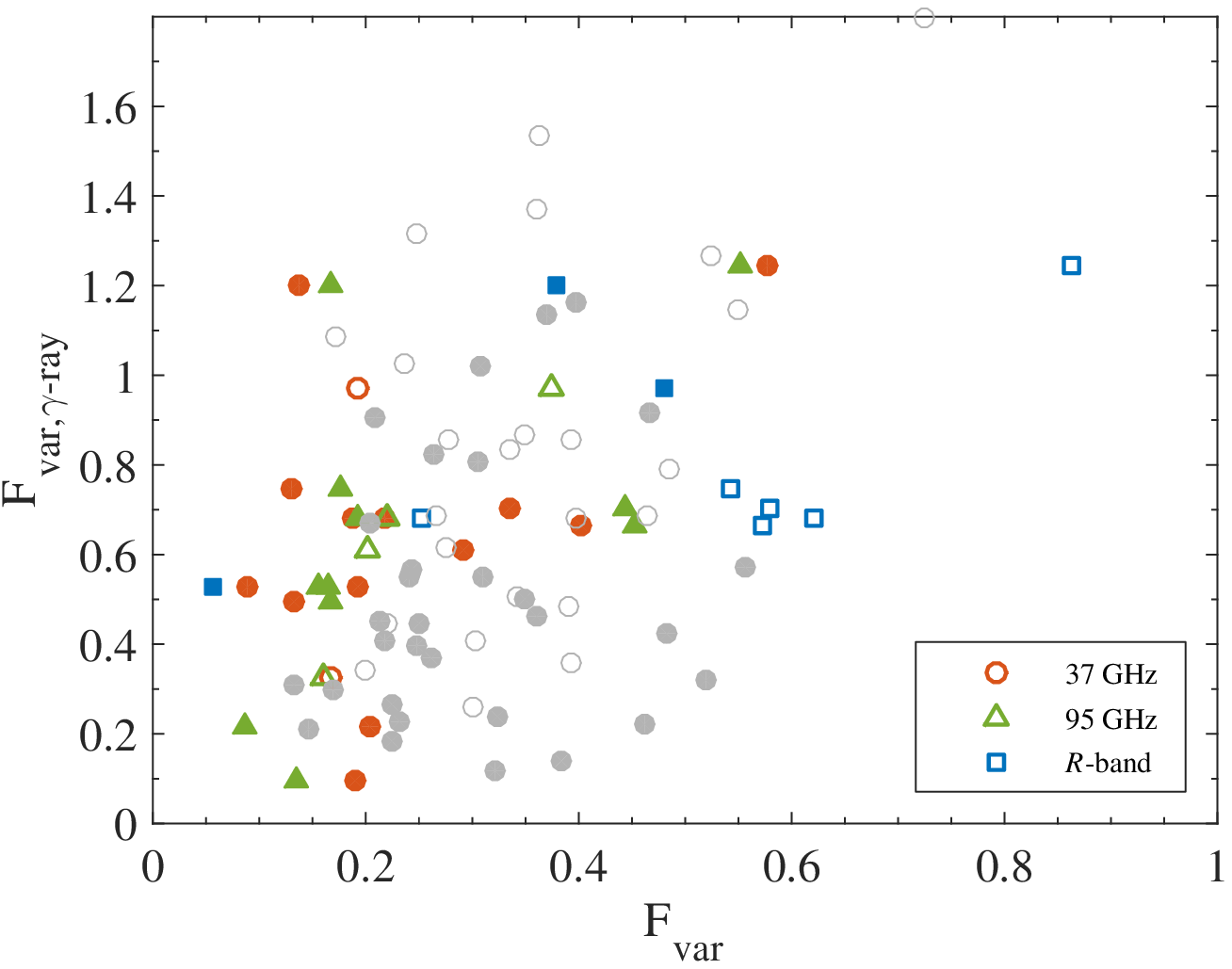}
	\includegraphics[width=\columnwidth]{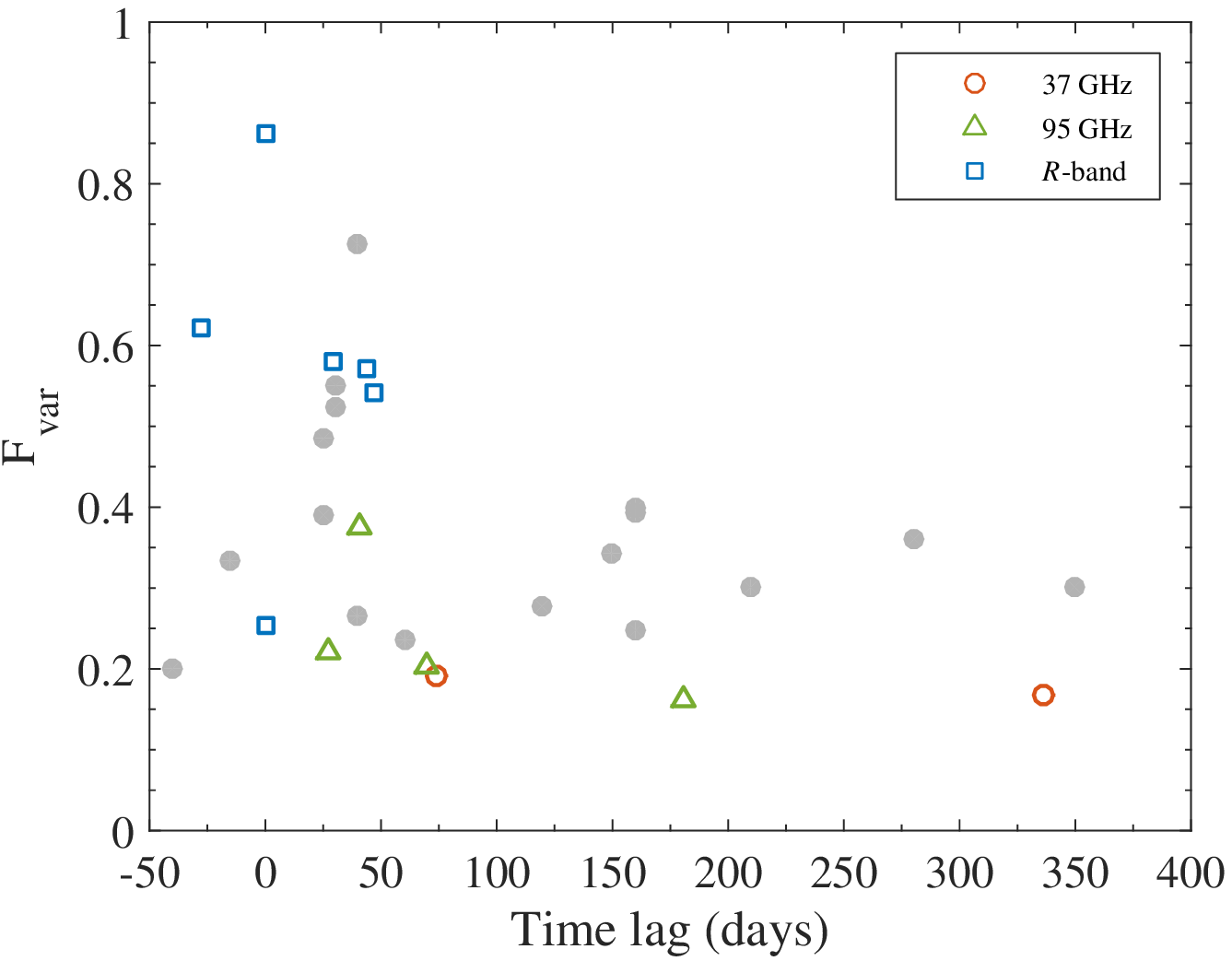}
	\includegraphics[width=0.5\textwidth]{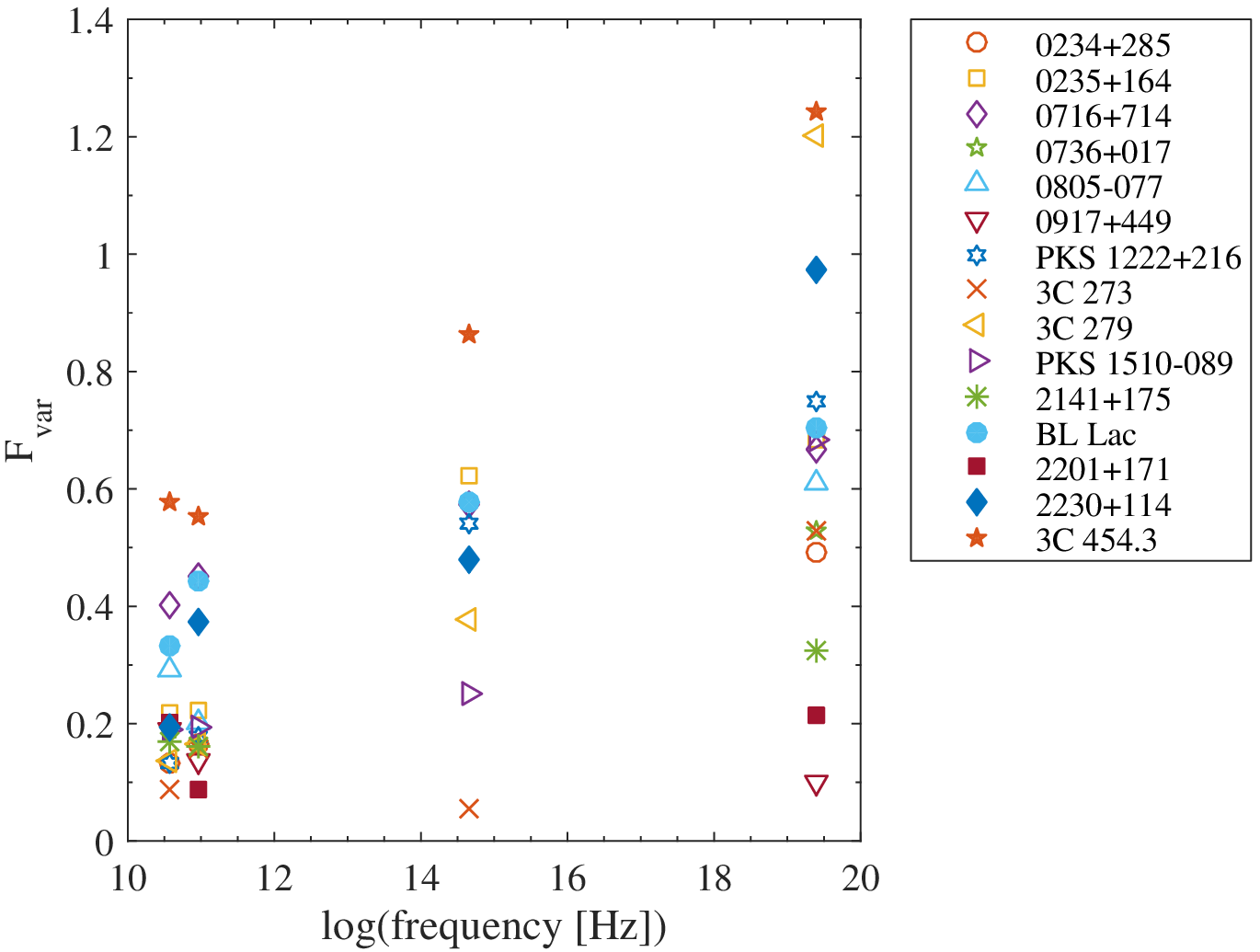}
	\caption{Top-Left: \fvar of correlated (open symbols) and uncorrelated (filled) sources at millimetre and \textit{R}-band versus those at $\gamma$ rays. In gray are the \fvar of 55 sources from Paper~I. Top-Right: Variation of fractional variability with time lag for sources with significant correlation in $\gamma$-ray vs. 37~GHz/95~GHz/\textit{R}-band (red/green/blue colours). The grey filled circles are \fvar of correlated sources from Paper~I. Bottom: Fractional variability as a function of frequency.}
	\label{fig2}
\end{figure*}

To quantify the variability of the sources at all wavebands we estimated the fractional root mean squared variability amplitude $F_{\rm var}$ from the relation \citep{Vaughan2003}
\begin{equation}
	F_{\rm var} = \sqrt{\frac{S^2 - \overline{\sigma^2_{\rm err}}}{\overline{x}^2}},
\end{equation}
where $S^2, \overline{\sigma^2_{\rm err}}$ and $\overline{x}^2$ are the sample variance, mean squared error and arithmetic mean of the observations, respectively. The uncertainty of $F_{\rm var}$ was determined as discussed in \citet{Vaughan2003}
\begin{equation}
	{\rm err}(F_{\rm var}) = \sqrt{ \left\{ \sqrt{\frac{1}{2N}} \frac{ \overline{\sigma_{\rm err}^{2}}}{ \bar{x}^{2}F_{\rm{var}} }  \right\}^{2} + 
		\left\{ \sqrt{\frac{\overline{\sigma_{\rm{err}}^{2}}}{N}}
			\frac{1}{\bar{x}}  \right\}^{2} }.
\end{equation}
where $N$ is the number of points in a light curve. The estimates along with their uncertainties are listed in Table~\ref{tab1}. The relation between \fvar in $\gamma$ rays and other wavebands is illustrated in the left-hand panel of Fig.~\ref{fig2}. No conclusive statement to discriminate the correlated sources from uncorrelated ones can be made from this analysis.

In the right-hand panel of Fig.~\ref{fig2} we show the variation of \fvar of the millimetre/optical bands versus time lags, along with that of the correlated sources from Paper~I. It is evident from Fig.~\ref{fig2} that sources with larger \fvar in the lower frequencies show shorter time lags. The \fvar as a function of frequency is shown in the bottom panel of Fig.~\ref{fig2}. From this figure the variability of sources can be seen to increase with frequency, except in source 3C~273 with an undersampled optical light curve and in 3C~454.3 which shows a minor decrease in \fvar at 95~GHz. This finding is in line with the previous studies on individual sources \citep[e.g.][]{Aleksic2015}.

\section{Discussion}
\label{sect7}

Visual inspection of the multifrequency light curves of these blazars shows that the optical and $\gamma$-ray variability are on daily or weekly time-scales, with longer time-scales of variability in the millimetre band. From the perspective of PSD slope, this corresponds to the slope steepening with decreasing frequency that can be understood from the values reported in Table~\ref{tab1}. To investigate the relation between variability at different bands, we performed cross-correlation analysis of 15 sources as described in Section~\ref{sect:dcf}. The time lags obtained from the cross-correlation analysis may be associated with the relative locations of the emission region at different wavebands.

We can summarize our findings as follows: (i) The 37 and 95~GHz variations are always correlated with a short time delay. (ii) If the optical light curve is well sampled, a significant correlation is seen between the optical and $\gamma$-ray variations. (iii) Most sources do not show significant correlations between the millimetre and optical/$\gamma$-ray bands. (iv) Some sources are faint or undersampled preventing us from characterizing the variability. Below we discuss all these cases and the possible reasons for the lack of correlation in more detail.

\subsection{Millimetre connection}

According to the prevailing theory of blazar emission, the emission in radio to optical (and sometimes also in X-rays) corresponds to synchrotron radiation from the jet. The variability observed in the millimetre band can be attributed to the shocks in the relativistic jet as explained by the shock-in-jet model \citep{Marscher1985,Hughes1985} and the frequency-dependent time delays can be explained by the generalized shock model \citep{Valtaoja1992a}.

The variations in the light curves at the millimetre bands studied in this work are simultaneous in most sources, resulting in a near-zero time delay in the correlation analysis. The only exceptions are sources 3C~273, 2141+175, and 2230+114, which show a time lag of $\sim$2~months with the 37~GHz lagging the high-frequency variations. This short delay and a clear one-to-one correspondence of the outbursts at both bands is also supported by the negligible difference in \fvar at 37 and 95~GHz (see Fig.~\ref{fig2}).

The short time delay between the two bands is somewhat surprising, considering the typical assumption that at 95~GHz the jets are closer to being optically thin, which should result in longer time delays to the 37~GHz emission. Our findings indicate that the opacity in these two frequencies is similar, at least in the majority of sources studied in this paper.

\subsection{Optical/$\gamma$-ray connection}
\label{opt_gamma}

The optical light curves of most sources show short time-scales in variability with large amplitude outbursts having a corresponding event in $\gamma$ rays. Interestingly, despite the fast variations present in the $\gamma$-ray light curve of 3C~273, no significant variability can be detected in optical. Regardless of the sparse sampling of the source in this work, a similar source state was also noted in \citet{Bonning2012}.

A time delay of one month or longer was obtained in 0716+714, 1222+216 and BL~Lac, with the $\gamma$ rays leading the optical variations. A zero-time delay was obtained in sources 1510$-$089 and 3C~454.3. The only source with the optical leading the $\gamma$-ray is 0235+164, with a time delay of 28~d. However, most of the correlations exhibit a broad peak with a width at half-maximum of about 100~d, preventing us from drawing firm conclusions about the delay. This broadness of the DCF peak could indicate the width of the respective flare in either band or the multiple correlation time-scales present. A much finer resolution in the light curve is required to investigate the delays of individual events. 

The correlation results obtained for optical and $\gamma$-ray light curves favour leptonic over hadronic models since no correlation is predicted by the latter \citep{Bottcher2007}. In \citet{CohenD2014} the authors reported that in FSRQs the $\gamma$-ray leads the optical by 0--20~d, suggesting EIC dominance in these sources while the SSC scenario was favoured in BLOs due to a near-zero time delay. Our result of the time delay in the range of 0--47~d in FSRQs is consistent with those obtained in \citet{CohenD2014}.

\subsection{Millimetre/$\gamma$-ray connection}
\label{mm_gamma}

It has been suggested that strong $\gamma$-ray detections occur during the rising phase or peak of the millimetre flare \citep[e.g.][]{Valtaoja1995,Lahteenmaki2003,Jonathan2011}. To investigate this connection further we studied the cross-correlation of 55 sources between the 37~GHz and $\gamma$-ray light curves in Paper~I. Based on 26 significant and stacked correlations we constrained the $\gamma$-ray emission region to the parsec-scale jet.

In contrast, in this work only two sources at 37~GHz and four sources at 95~GHz showed a correlation with the $\gamma$-ray band. All sources at 37~GHz in this work had the \fvar a factor of $\sim$2 smaller than that observed in Paper~I. This indicates that we did not observe as many large outbursts during the shorter time frame of this paper compared to Paper~I. Among well sampled sources 0234+285, 0235+164, 1222+216, and 3C~273 fall into this category. None of them showed as large amplitude outbursts as those observed in Paper~I.

This is not surprising considering that the variability time-scales in these sources at the millimetre bands are long with large flares occurring on average every four to six years \citep{Hovatta2007}. Therefore, in the 2.5~yr time span studied in this paper, we expect less large outbursts than during the 5~yrs used in Paper~I. To validate this argument on the significance of the cross-correlation, we extended the time span of the 37~GHz light curve to 2015 February for the source 0235+164. By doing so we included the rising flux that was already noticed in the 95~GHz light curve. We then ran the cross-correlations for the new 37~GHz data and found a significant correlation with the 37~GHz lagging the $\gamma$ rays by $\sim$200~d. This test was performed only to show the effect of the shorter time span on the correlations and is not reported in Table~\ref{tab2}.

On the other hand, sources 0716+714, 0736+017, 0805$-$077, 1510$-$089, and 2230+114 show persistent variability both in $\gamma$ rays and millimetre band even during the time span considered here. Visual inspection of these sources shows a marginal connection of the events at both these bands which is also visible in the DCF, although the peaks are not significant at 99~per cent level in many cases. None of these sources, except 0805$-$077 showed a significant correlation in Paper~I.

Blazars such as 3C~279, BL~Lac, and 3C~454.3 that are bright and highly variable in all the bands, show an increasing or decreasing trend in the millimetre light curves. None of these sources showed a significant correlation in the millimetre and $\gamma$-ray analysis. This could be affected by the flare evolution that is studied in this work; only the peak or decay part in the case of 3C~279 and BL~Lac, and the rising part of the flare for 3C~454.3 is visible in the millimetre light curves. This indicates that either the time delays between the $\gamma$-ray and millimetre variations in these sources are longer than can be probed with our light curves, or that the smaller amplitude variations are not correlated in these bands.

Our paper includes three objects, 0917+449, 2141+175, and 2201+171 that are faint in both millimetre and $\gamma$-ray bands and show little variability. Of these 2141+175 is the only source with a modest number of $\gamma$-ray detections, and it also shows a significant correlation in millimetre and $\gamma$-ray analysis. Neither 0917+449 nor 2201+171 was reported with a correlation at a time lag <400~d in Paper~I.

\subsection{Millimetre/optical connection}

Blazars usually exhibit more flares with shorter variability time-scales in the optical than in the millimetre bands \citep[e.g.][]{Hufnagel1992,Tornikoski1994}. The occurrence of optical and radio flares, in general, can either be simultaneous or with a time delay that increases towards lower frequencies. However, due to the uneven sampling and the seasonal gaps in the optical light curves many events are missed. The correlations are also affected by the millimetre light curves consisting of superposed events.

The correlations between the optical and radio emission have been reported in \citet{Tornikoski1994} and \citet{Arshakian2010b}. The localization of the optical emission region to the parsec-scale jet and the emission process being non-thermal was discussed by \citet{Arshakian2010a} in 3C~390.3 and \citet{Jonathan2010} in 3C~120, based on the connection between the optical continuum variability and kinematics of the jet. Our analysis showed only two sources in 37~GHz and three in 95~GHz with a strong correlation with the optical light curves. Apart from these sources, 1510-089 shows a moderate correlation with a time delay at $\sim$50~d with the optical leading the millimetre bands. Our millimetre to optical correlations, therefore, agree with the results presented above where no significant correlations were found between the millimetre and $\gamma$-ray bands, while optical variations were found to be significantly correlated with the $\gamma$ rays.

\section{Summary}
\label{sect8}
We extend the study of constraining the $\gamma$-ray emission site performed in Paper~I through cross-correlation analysis of multifrequency light curves in 15 blazars over the interval 2012.6--2014.9. The time lags from the correlations were obtained using DCF with significance level at 99~per cent. Overall the time lags obtained from all cross-correlations show a decrease with increasing frequency in accord with the synchrotron self-absorption effects. The \fvar estimated was found to increase with frequency. A shorter time lag was obtained in sources with significant correlations with higher \fvar in millimetre and optical bands, versus the $\gamma$-ray emission.

Our results in the millimetre band show a near-zero time delay for most of the sources. This indicates that for correlation studies between the millimetre and other energy domains the 37~GHz data are in most cases as useful as higher-frequency radio data, which are typically harder to obtain due to the lack of dedicated, well-sampled long-term monitoring programmes.

The optical and $\gamma$-ray correlation was strong in six sources: three sources with optical lagging the $\gamma$ rays by a month or longer, two with zero-time delay, and one with optical leading the $\gamma$ rays by a month. In all six sources the optical light curve was well sampled. The presence of significant correlation supports the leptonic model for the $\gamma$-ray emission.

A correlation between the millimetre and $\gamma$-ray data is significant in two and four sources detected in 37~GHz and 95~GHz bands, respectively. From visual inspection of the millimetre and $\gamma$-ray light curves we find that the following reasons may explain the lack of correlation in the rest of the sample:

\begin{itemize}
	\item[(i)] shorter time span -- within the 2.5~yr time frame considered in this work not many sources are expected to display a large outburst, since their chances of occurrence are on average four to six years in the millimetre band;
	\item[(ii)] moderate variability -- the modest variations in $\gamma$ rays and at both millimetre bands in some sources makes the DCF peak less pronounced;
	\item[(iii)] large events -- the lack of significant correlation in three bright blazars could possibly be due to the increasing or decreasing trend in their millimetre light curves, which could possibly result in longer time delays, i.e. beyond the time span studied in this work;
	\item[(iv)] faint sources -- three objects are faint both in millimetre and $\gamma$-ray bands with only one among them showing a significant correlation.
\end{itemize}

Three sources showed a significant correlation between the 95~GHz and optical band, and two of them also between the 37~GHz and optical band. The lack of significant correlations in this analysis is partly affected by the undersampled optical data and the presence of seasonal gaps.

Thus, the lack of significant correlations in most cases restricts us from placing stringent limits on the $\gamma$-ray emission region and, therefore, the physical mechanism responsible for the $\gamma$-ray emission in these sources.

\section*{ACKNOWLEDGEMENTS}
We thank the anonymous referee for valuable comments that improved the manuscript. VR acknowledges the support from the Finnish Graduate School in Astronomy and Space Physics. TH was supported by the Academy of Finland project number 267324. MB acknowledges support from the International Fulbright Science and Technology Award, and the NASA Earth and Space Science Fellowship Program, grant NNX14AQ07H. We are grateful for the support from the Academy of Finland to the Mets\"ahovi AGN monitoring project (numbers 212656, 210338, 121148 and others). Support for CARMA construction was derived from the states of California, Illinois, and Maryland, the James S. McDonnell Foundation, the Gordon and Betty Moore Foundation, the Kenneth T. and Eileen L. Norris Foundation, the University of Chicago, the Associates of the California Institute of Technology, and the National Science Foundation. Ongoing CARMA development and operations are supported by the National Science Foundation under a cooperative agreement, and by the CARMA partner universities. Data from the Steward Observatory spectropolarimetric monitoring project were used. This program is supported by \textit{Fermi} Guest Investigator grants NNX08AW56G, NNX09AU10G, and NNX12AO93G. This paper has made use of up-to-date SMARTS optical/near-infrared light curves . We also acknowledge the computational resources provided by the Aalto Science-IT project. This research has made use of NASA's Astrophysics Data System.

\bibliographystyle{mnras}
\bibliography{mfcorr}

\appendix

\section{Notes on Individual sources}

\textit{0234+285}: The multifrequency light curve of the source is characterized by short-term variations with decreasing trend from the major flare that occurred during early 2012 in the millimetre band, and one flare in the $\gamma$-ray band. The $\gamma$-ray flare was by a factor of 3 smaller than the one that occurred toward the end of 2011 (see Paper~I). No significant correlations between the $\gamma$-ray and millimetre bands were obtained, unlike the 60~d lag reported in Paper~I. The cross-correlations within the millimetre bands were significant with a time lag of 6~d with the 95~GHz preceding the 37~GHz light curve.

\textit{0235+164}: The quasi-simultaneous variability in the 95~GHz, \textit{R}-band, and $\gamma$-ray light curves towards the end of the year 2014 in this source, yielded the 95~GHz lagging the $\gamma$ rays by 27~d and the \textit{R}-band preceding the $\gamma$ rays by 28~d. No significant correlation was obtained between 37~GHz and $\gamma$ rays. In Paper~I we reported a correlation with $\gamma$ rays leading by 30~d. 

\textit{0716+714}: The persistent variability of the source at all wavebands manifests in the DCF with a pronounced peak although not being at 99~per cent level in all cases. Our correlation results showed variations in the millimetre bands, and those between the 95~GHz and \textit{R}-band, and $\gamma$ rays and \textit{R}-band to be correlated at >99~per cent significance level. Many authors have investigated the multifrequency behaviour of this source \citep[e.g.][]{Larionov2013,Rani2013,Rani2014}, due to its rapid variability across the electromagnetic spectrum. The time delay of 44~d obtained between optical and $\gamma$ rays, can be related to the zero-time delay reported in \citet{Larionov2013} and \citet{Rani2013} since the DCF peak is broad and also contains zero-time delay.

\textit{0736+017}: The $\gamma$-ray data in this source is characterized by sporadic low-amplitude variations while it is less variable in the millimetre band. Despite the coincidence of the high-state around 2014.5 in millimetre and $\gamma$ rays, a correlation only at 95~per cent significance level with a time delay of 360~d was obtained.

\textit{0805-077}: The activity of this source in $\gamma$ rays is characterized by continuous variability with one major activity period around 2013.5. A high-intensity outburst around 2009.5 with strong variability throughout that year was observed in Paper~I. Although the 2013.5 activity is not the brightest feature observed in the $\gamma$ rays, it yields a significant correlation with the 95~GHz light curve with the 95~GHz lagging the $\gamma$ ray variations by 70~d. Despite the different time window in Paper~I, the time lag of 120~d with the 37~GHz emission lagging the $\gamma$ rays seems to be in agreement with this result, given the higher observing frequency.

\textit{0917+449}: This FSRQ was active during the first two years of \textit{Fermi}/LAT operation and since then has showed no pronounced activity in $\gamma$ rays. The cross-correlation in the millimetre waveband was significant at a time lag $-$8~d with the 95~GHz lagging the 37~GHz variations. The $\gamma$-ray detections occur mostly during the rise of the millimetre flare. However, no correlation with the $\gamma$ rays was observed with the light curve being dominated by its noise.

\textit{1222+216}: The $\gamma$-ray variations for the time interval in this work are much smaller than those observed in Paper~I. In this source, the $\gamma$-ray flare around 2014.2 corresponds closely in time to the \textit{R}-band flare with a time lag of 47~d. No other correlation with the $\gamma$ rays was obtained. The \textit{R}-band, however, resulted in a significant correlation with the 95~GHz emission at a time lag of 261~d. The correlated variation in the millimetre was such that the 37~GHz lead the 95~GHz by 6~d.

\textit{3C~273}: In $\gamma$ rays 3C~273 was active for almost two years since \textit{Fermi}/LAT began its operation. Since then no exceptional flaring has been reported at this frequency. In this work, a small amplitude flare around 2014.7 was observed in the $\gamma$ rays. No significant correlation was obtained between the light curve in $\gamma$-ray band and those at other wavebands. In the millimetre bands a flare lasting for almost a year is seen in 2013. The correlation result shows the 37~GHz lagging the 95~GHz emission by 79~d.

\textit{3C~279}: The $\gamma$-ray flare that occurred on 2014 April 03 was the brightest event observed at this band \citep{Hayashida2015}. Despite the coincidence of the $\gamma$-ray events in 3C~279 to the optical variations, no significant correlation was obtained. Neither did we obtain a significant correlation between other wavebands. The millimetre light curves needed to be detrended prior to the correlations due to the decreasing trend from the decay of the major flare that occurred toward the middle of 2012.

\textit{1510$-$089}: The $\gamma$-ray variations in this source are by a factor $>2$ larger in Paper~I compared to those analysed here. The $\gamma$-ray and the optical flare around 2013.7 correlated with a zero-time lag. In the millimetre waveband the 37~GHz lagged those at 95~GHz by 5~d, implying the variability at both these bands to be synchronous.

\textit{2141+175}: The FSRQ 2141+175 portrays sporadic variations in the $\gamma$ rays with one marginal activity period around 2013.3. In the millimetre band a flare around 2012.6 was noticed. A time lag of 181 and 336~d was obtained with the 95 and 37~GHz variations lagging those at $\gamma$ rays, respectively. In millimetre bands, the 37~GHz lagged the 95~GHz ones by 65~d.

\textit{BL~Lac}: In the case of BL~Lac the variability in $\gamma$ rays and \textit{R}-band are in unison, and so are the variations within the millimetre wavebands. The $\gamma$-ray flare around 2012.6 comes from a series of outbursts that began by 2011 (see Paper~I). In the millimetre band the largest recorded event can be observed during 2012.9. Owing to the decreasing trend in this band, the light curves were detrended prior to the correlations. These light curves showed no significant correlation with neither the \textit{R}-band nor the $\gamma$ rays, due to the distinct periodic variations in the detrended light curves. The correlation between the millimetre bands was, however, detected at >99~per cent significance level at a time lag of 7~d. At the higher frequencies the activity period around 2013.9 observed both in the $\gamma$ rays and \textit{R}-band, coupled with the high state around 2012.5, yielded a correlation at a time lag of 29~d with the emission in \textit{R}-band lagging. The optical light curve of this source both from KVA and Steward Observatory were coincident in the low state. However, during the outburst around 2013.9, significant differences were noticed between the light curves from the two observatories. We investigated this by correlating the light curve with the $\gamma$-ray, both including and excluding the Steward data, and conclude that the difference in the correlation result was negligible. Hence, the Steward observations were also included in the analysis. This, however, affected the PSD analysis with the slope changing from 0.99 to 1.27 when Steward observations were not included, despite not affecting the significance level in correlations. Hence the correlation result reported here is not sensitive to the change in the PSD slope.

\textit{2201+171}: This fairly faint source in all the bands showed a year-long activity periods in the millimetre data. The 37~GHz variations lagged those at 95~GHz with a time delay of 12~d. In $\gamma$ rays we had a few detections that were insufficient for producing any significant correlation.

\textit{2230+114}: Strong variability can be observed across the wavebands in this source. The $\gamma$-ray event around 2012.7 was the largest recorded in this source. Among all the correlated sources this is the only one with significant correlation in all the cross-correlations, except those at \textit{R}-band versus the $\gamma$ rays which could be due to the undersampled \textit{R}-band data. The time lags are in the range of 41--117~d, with the lower frequency always lagging those at higher. The time delay of 41~d obtained in the 95~GHz versus $\gamma$-ray correlation here is consistent with the finding in \citet{Casadio2015}, where the authors report a shift in the peak of the 1~mm light curve by about a month when compared to the $\gamma$-ray peak around 2012.7. The cross-correlation analysis of \citet{Casadio2015} showed a zero-time delay in optical and $\gamma$-ray analysis. Meanwhile, in the millimetre and $\gamma$-ray analysis no significant correlation was obtained in their work, which could be due to the shorter time span (about 200~d), considered for the cross-correlation study.

\textit{3C~454.3}: The FSRQ 3C~454.3 is one of the most widely studied sources. Small-scale variations in amplitude characterize the variability of the source in the $\gamma$ rays until the middle of 2014, followed by an active phase for over 6 months. The zero-time lag between the $\gamma$ rays and \textit{R}-band indicates simultaneous variability in these wavebands. This time delay is in good agreement with results obtained by others \citep{Bonning2009,Jorstad2013}. The light curves in millimetre bands show an increasing trend. Hence, we correlated the detrended millimetre light curves with other wavebands. The detrended 37~GHz light curve correlated with the \textit{R}-band at a time lag of 272~d. No distinct association of activity in the millimetre and higher-frequency light curves could be firmly established prior to removing the trend.

\label{lastpage}

\end{document}